\documentclass[prd,preprint,tightenlines,floatfix,showpacs,preprintnumbers,nofootinbib,eqsecnum]{revtex4}
\usepackage[dvips,final]{graphicx}
  \usepackage{amssymb}
   \usepackage{amsmath}
    \usepackage{amsfonts}
     \usepackage{epsfig}
      \usepackage{bm}

\begin{document}

\thispagestyle{empty} \preprint{\hbox{}} \vspace*{-10mm}

\title{Factorization method in the model of unstable particles with a smeared mass}

\author{V.~I.~Kuksa}

\email{kuksa@list.ru}

\affiliation{Institute of Physics,
Southern Federal University ,
Rostov-on-Don, Russia}

\date{\today}

\begin{abstract}
The method of factorization, based on the model of unstable particles with a smeared mass,
is applied to the processes with an unstable particle in the intermediate state. It
was shown, that in the framework of the method suggested, the decay rate and cross-section
can be represented in the universal factorized form for an
arbitrary set of particles. An exact factorization is caused by the specific
structure of unstable particles propagators. We performed the phenomenological analysis of the
factorization effect.
\end{abstract}


\maketitle

\section{Introduction}

The most of the elementary (fundamental) particles are unstable,
however, a large width have the $W, Z$ bosons and $t$ quark only. For the rest of unstable particles
the ratio $\Gamma/M$ is very small, therefore the so-called stable particle approximation is valid with very
high precision. The most of hadrons (mesons and baryons) have a large width and this approximation is not valid.
So, we have to take into account the finite-width effects (or instability) in the processes with the participation of the unstable particles or resonances with a large width.

The specific properties of the unstable particles (UP) were being under considerable
discussion during the last decades.
In particular, the assumption that the decay of UP or resonance (R) proceeds
independently of its production remains of interest \cite{1,2,3}.
Formally, this effect is expressed as the factorization of a
cross-section or decay rate \cite{3}. The processes of type
$ab\rightarrow Rx\rightarrow cdx$ were considered in
Ref.\cite{3}. It was shown, that the factorization always is valid
for a scalar $R$ and does not take place for a vector and spinor $R$.
The factorization usually is related with the narrow-width
approximation (NWA) \cite{4}, which makes five critical
assumptions \cite{5}.

We consider the factorization method, which is based on the model of UP with a smeared mass \cite{6,6a}
and related with the propagator structure. The decay processes of type $a\rightarrow Rx\rightarrow cdx$
were analyzed in Ref.\cite{7}. It was shown in this work, that the factorization always is valid
for a scalar $R$, while for a vector and spinor $R$ it occurs when the propagators' numerators are
$\eta_{\mu\nu}(q)=g_{\mu\nu}-q_{\mu}q_{\nu}/q^2$ and $\hat{\eta}(q)=\hat{q}+q$, respectively, where
$\hat{q}=q_i\gamma^i$ and $q=\sqrt{q_iq^i}$. The processes of type
$ab\rightarrow R\rightarrow cd$, were considered in Ref.\cite{8}. It was shown,
that the cross-section $\sigma(ab\rightarrow R\rightarrow cd)$ can be
represented in the universal factorized form when the same expressions
$\eta_{\mu\nu}(q)$ and $\hat{\eta}(q)$ are used to describe the propagator's
numerator of vector and spinor UP. Such a structure of
propagators always provides an exact factorization for any tree
process. This condition of factorization is some analytical analog of NWA,
which is discussed in Section 3 and 4.
These propagators were constructed in the model of UP with a smeared mass \cite{6,6a} and
describe some effective (dressed by self-energy insertion) unstable fields in an intermediate state.
The model have been applied in the various fields of particles physics \cite{6,6a}.
Note that the structure of the expressions $\eta_{\mu\nu}(q)$ and $\hat{\eta}(q)$
is not related with the choice of the gauge (see the second section).

In this work, we systematically analyze the effects of factorization in the processes with UP in an intermediate state. In Section 2 we illustrate the premise of factorization and give universal factorized formulae for the decay rate of three-particle decay and for the cross-section of two-particle scattering. The factorization approach is applied to the processes of scattering with consequent decays of the final states (Section 3). It was noted, that similar processes were considered in Refs.\cite{8a,8b}, where the phenomenon of pseudoresonances was discussed. In Section 4, we analyze some methodological and phenomenological
aspects of factorization.

\section{Factorization effect in the model of unstable particles with a smeared mass}

In this section, we consider the structure of the model amplitude when UP is in the intermediate state. We show that the special form of the model propagators of unstable fields lead to the factorization of the transition probability. In contrast to the traditional treatment (narrow-width approximation, NWA), the approach suggested provides an exact factorization for the any type of UP. This effect makes it possible to represent the decay rate of three-particle decays and the cross-section of two-particle scattering in the universal factorized form.

The model propagators of scalar, vector and spinor unstable fields are defined by the following expressions (see Appendix 1):
\begin{equation}\label{2.1}
 \frac{i}{P(q^2)};\,\,\,-i\frac{g_{\mu\nu}-q_{\mu}q_{\nu}/q^2}{P(q^2)};\,\,\,i\frac{\hat{q}+q}{P(q^2)}.
\end{equation}
In Eqs.(\ref{2.1}) $q^2=(q_iq^i)$, $q=\sqrt{(q_iq^i)}$ and $P(q^2)$ can be defined in arbitrary way (pole, Breit-Wigner and other definitions). It is essential, that the effect of factorization does not depend on the definition of denominator $P(q^2)$ and crucially depends on the structure of propagator's numerators for the case of vector and spinor fields. The model expressions $\eta_{\mu\nu}(q)=g_{\mu\nu}-q_{\mu}q_{\nu}/q^2$ and $\hat{\eta}(q)=\hat{q}+q$ provide an exact factorization, while the traditional expressions $\eta_{\mu\nu}(M)=g_{\mu\nu}-q_{\mu}q_{\nu}/M^2$ and $\hat{\eta}(M)=\hat{q}+M$ leads to an approximate factorization (NWA). It should be noted, that the structure of $\eta_{\mu\nu}(q)$ and $\hat{\eta}(q)$ is not related with the choice of the gauge. The model under consideration is not a gauge one and describes some effective unstable fields (see Appendix 1). We note, also, that the differences between the model and traditional $\eta$ -functions are small at $q^2\approx M^2$. So, the model approach can be treated as approximation to the standard one, that is gives an analytical alternative of NWA (see Section 4)

Now, we consider the mechanism of factorization in the processes of three-particle decay $\Phi_1\to\Phi_2 R\to \Phi_2\Phi_3\Phi_4$ and two-particle scattering $\Phi_1\Phi_2\to R\to \Phi_3\Phi_4$. In the case of vector UP in an intermediate state the model amplitude is
\begin{equation}\label{2.2}
 \mathit{M}\sim \Phi_1\Gamma^{\mu} \Phi_2 \frac{g_{\mu\nu}-q_{\mu}q_{\nu}/q^2}{P(q^2)}\Phi_3\Gamma^{\nu}\Phi_4.
\end{equation}
 It is essential, that the structure of propagator's numerator and polarization matrix is the same (Appendix 1):
\begin{equation}\label{2.3}
 \sum_{k=1}^{3} e^k_{\mu}(q) e^{*k}_{\nu}(q)=-(g_{\mu\nu}-q_{\mu}q_{\nu}/q^2).
\end{equation}
 Thus, from Eqs.(\ref{2.2}) and (\ref{2.3}) it follows
\begin{equation}\label{2.4}
 \mathit{M}\sim \sum_{k=1}^{3}\frac{\mathit{M}_{1}^{(k)}\cdot\mathit{M}_{2}^{(k)}}{P(q^2)}\,,
\end{equation}
where $\mathit{M}_{1}^{(k)}\sim\Phi_1\Gamma^{\mu}\Phi_2 e^k_{\mu}$ and $\mathit{M}_{2}^{(k)}\sim\Phi_3\Gamma^{\nu}\Phi_4 e^k_{\nu}$. From Eq.(\ref{2.4}) it follows, that for the case of scalar UP, exact factorization occurs at amplitude level (see also \cite{3,7}).
The quasifactorized structure of the full amplitude $\mathit{M}$ is direct consequence of Eq.(\ref{2.3}), that is of the smearing of mass-shell. Full factorization occur in the $|\mathit{M}|^2$, when the properties of the polarization matrixes of the initial and final states are used. In the standard treatment factorization takes place when the intermediate state occur on mass-shell $q^2=M^2$, while the virtual states destroy the factorization. In our approach this effect takes place at arbitrary $q^2$ due to smearing (fuzzing) of mass-shell (see Appendix 1) and some dualism of virtual and real states. This dualism imply the possibility to describe UP by polarization matrix (real state) and propagators (virtual state) at the same time \cite{6,7}. More exactly, the division of the unstable states onto virtual and real ones have no sense in the vicinity of the resonance.

The same effect takes place for the case of spinor UP in an intermediate state. In this case, the structure of spinor propagator's numerator $\hat{\eta}(q)=\hat{q}+q$ is similar to the structure of spinor polarization matrix (Appendix 1):
\begin{equation}\label{2.5}
 \sum_{a=1}^{2} u^{a,\pm}_{\alpha}(q) \bar{u}^{a,\mp}_{\beta}(q)=\frac{1}{2q^0}(\hat{q}\mp q)_{\alpha\beta}.
\end{equation}
Thus, the premise of factorization is the coincidence of the polarization matrix and propagator's numerator for any $q^2$, which is directly related with the smearing (fuzzing) of mass-shell.

Let us consider the three-particle decay of type $\Phi\to \phi_1 R\to \phi_1\phi_2\phi_3$, where $R$ is UP of any kind with a large width. The method of calculation and some specific details of the model approach are given in Appendix 2. By straightforward calculation it was checked, that the decay rates of the processes under consideration can be represented in the universal factorized form:
\begin{equation}\label{2.6}
  \Gamma(\Phi\rightarrow\phi_1 \phi_2 \phi_3)=\int_{q^2_1}^{q^2_2}\Gamma(\Phi
  \rightarrow\phi_1 R(q))\frac{q\Gamma(R(q)\rightarrow\phi_2\phi_3)}
  {\pi \vert P_{R}(q)\vert ^2}dq^2\,,
\end{equation}
where $q_1=m_2+m_3$ and $q_2=m_{\Phi}-m_1$. By means of the summation over decay channels of $R$, from Eq.(\ref{2.6}) we get the well-known convolution formula for the decays with UP in a final state \cite{6a,7,9,10}:
\begin{equation}\label{2.7}
 \Gamma(\Phi\rightarrow\phi_1 R) = \int_{q^2_1}^{q^2_2} \Gamma(\Phi\rightarrow
 \phi_1 R(q))\rho_{R}(q)dq^2\,.
\end{equation}
In Eq.(\ref{2.7}) the smearing of mass of unstable state $R$ is described by the probability density $\rho_R(q)$:
\begin{equation}\label{2.8}
 \rho_{R}(q)=\frac{q\Gamma^{tot}_R(q)}{\pi\vert P_{R}(q)\vert^2}.
\end{equation}
If the parametrization $q\Gamma(q)=Im\Sigma(q)$ and
Dyson-resummed propagator are used, then we get:
\begin{equation}\label{2.9}
 P_{R}(q)=q^2-m^2_{R}(q)-iIm\Sigma_{R}(q),\,\,
 m^2_{R}(q)=m^2_{0R}+Re\Sigma_{R}(q),
\end{equation}
and the $\rho_{R}(q)$ can be written in the Lorentzian
(Breit-Wigner type) form:
\begin{equation}\label{2.10}
 \rho_{R}(q)=\frac{1}{\pi}\,\frac{Im\Sigma_{R}(q)}{[q^2-m^2_{R}(q)]^2+[Im\Sigma_{
 R}(q)]^2}\,.
\end{equation}
The expressions similar to (\ref{2.10}) have been used in the many papers \cite{6}-\cite{10}.

Now, we consider the two-particle scattering of type $a+b\to R\to c+d$, where $R$ is UP with a large width.
With the help of the expressions (\ref{2.1})
we have got by straightforward calculations (see Appendix 2) the
universal factorized formula for the cross-section for all permissible
combinations of particles $(a,b,R,c,d)$:
\begin{equation}
 \sigma(ab\rightarrow R\rightarrow cd)=\frac{16\pi (2J_R+1)}
 {(2J_a+1)(2J_b+1) \bar{\lambda}^2(m_a,m_b;\sqrt{s})}
 \frac{\Gamma^{ab}_R(s)\Gamma^{cd}_R(s)}{|P_R(s)|^2}.
\label{2.11}
\end{equation}
In Eq.(\ref{2.11})  $J_k$ is spin of the particle ($k=a,b,R$),
$s=(p_1+p_2)^2$, $\Gamma^{ab}_R(s)=\Gamma(R(s)\rightarrow ab)$ and
$P_R(s)$ is propagator's denominator of the UP or resonance $R$.
The expressions for $\Gamma^{ab}_R(s)$ and $\Gamma^{cd}_R(s)$
follow from the standard ones (see Appendix 2), when squared mass of UP
is $m^2_R=q^2=s$. The factorization of cross-section
does not depend on the definition of $P_R(s)$, which can be
determined in a phenomenological way, in Breit-Wigner or pole form
etc. The expression (\ref{2.11}) is a natural
generalization of the spin-averaged Breit-Wigner (non-relativistic)
cross-section, defined by the expression (37.51) in Ref.
\cite{11}. Note that the factorization is exact in our approach,
while in the traditional one it occurs as an approximation.

The cross-section of exclusive process $ab\rightarrow
R\rightarrow cd$, defined by Eq.(\ref{2.11}), does not depend on
$J_c$ and $J_d$. So, it can be summarized over final channels
$R\rightarrow cd$:
\begin{equation}
 \sigma(ab\rightarrow R(s)\to all)=\frac{16\pi k_R}
 {k_a k_b\bar{\lambda}^2(m_a,m_b;\sqrt{s})}\frac{\Gamma^{ab}_R(s)\Gamma^{tot}_R(s)}
 {|P_R(s)|^2}.
 \label{2.12}
\end{equation}
In Eq.(\ref{2.12}) $k_i=2J_i+1$ and
$\Gamma^{tot}_R(s)=\sum_{cd}\Gamma^{cd}_R(s)$, where for
simplicity we restrict ourselves by two-particle channels.

The factorization effect, expressed by Eq.(\ref{2.11}), has two
aspects. On the one hand, it means that the decay of UP proceeds
independently of its production in the approach considered. On the
other hand, it leads to the significant simplification of calculations,
in particular, in the case of the complicated processes
(see the next section).

\section{Factorization effect in the complicated processes}
In this section, we consider the factorization effects in the case of complicated chain processes.
For example, let us consider the decay-chain process $\Phi\to aR\to abR_1\to abcd$. It is convenient
to divide this process onto the stages $\Phi\to aR\to abR_1$ and $R_1\to cd$. In according with the
Eq.(\ref{2.6}) the width of the first process is
\begin{equation}\label{3.1}
  \Gamma(\Phi\rightarrow abR_1)=\int_{q^2_1}^{q^2_2}\Gamma(\Phi
  \rightarrow a R(q))\frac{q\Gamma(R(q)\rightarrow bR_1)}
  {\pi \vert P_{R}(q)\vert ^2}\,dq^2\,,
\end{equation}
where $\Gamma(R(q)\rightarrow bR_1)$ includes all decay channels of $R_1$. Analogously, the width of
exclusive decay $R(q)\to bR_1\to bcd$ is defined by the expression:
\begin{equation}\label{3.2}
  \Gamma(R(q)\rightarrow bcd)=\int_{g^2_1}^{g^2_2}\Gamma(R(q)\to bR_1(g))
  \frac{g\Gamma(R_1(g)\rightarrow cd)}
  {\pi \vert P_{R_1}(g)\vert ^2}\,dg^2\,,
\end{equation}
where $g_1=m_c+m_d$ and $g_2=m_{\Phi}-m_a-m_b$. Combining the expressions (\ref{3.1}) and (\ref{3.2}),
we get the width of
the full decay-chain process:
\begin{equation}\label{3.3}
\Gamma(\Phi\to abcd)=\frac{1}{\pi^2}\int_ {q^2_1}^{q^2_2}\frac{q\Gamma(\Phi
  \rightarrow a R(q))}{\vert P_{R}(q)\vert ^2}\int_{g^2_1}^{g^2_2}\Gamma(R(q)\to bR_1(g))
  \frac{g\Gamma(R_1(g)\rightarrow cd)}
  {\vert P_{R_1}(g)\vert ^2}dg^2\,dq^2.
\end{equation}
Using this method, one can write the width for the more complicated decay-chain processes. We should note that the
factorization reduces the number of independent kinematical variables which specify a point in the phase space.
In the general case of $n$-particle decay the number of such variable is $N=3n-7$ \cite{20}. Thus, in the standard
approach, for three- and four-particle decays we have $N=2$ and $N=5$. The factorization effect reduces these numbers
and gives $N=1$ (Eq.(\ref{2.6})) and $N=2$ (Eq.(\ref{3.3})), respectively.

Now, we consider the scattering $ab\to R\to x R_1$ with consequent decay $R_1\to cd$. In this case, Eq.(\ref{2.11})
has the form:
\begin{equation}
 \sigma(ab\rightarrow R(s)\to x R_1)=\frac{16\pi k_R}
 {k_a k_b\bar{\lambda}^2(m_a,m_b;\sqrt{s})}\frac{\Gamma^{ab}_R(s)\Gamma^{R_1x}_R(s)}
 {|P_R(s)|^2}.
 \label{3.4}
\end{equation}
 To calculate the value $\Gamma^{R_1 x}_R (s)$ we apply the convolution formula (\ref{2.7}), which accounts
 FWE in the decay $R(s)\to x R_1$:
\begin{equation}
 \Gamma(R(s)\to x R_1)=\int_{q^2_1}^{q^2_2}\Gamma(R(s)\to x R_1(q))\rho_{R_1}(q)\,dq^2\,.
 \label{3.5}
\end{equation}
In Eq.(\ref{3.5}) $q=p_R-p_x$, $q_{1,2}$ are defined by kinematics of the process and
$\rho_{R_1}(q)=q \Gamma^{tot}_{R_1}(q)/\pi|P_{R_1}(q)|^2$ is interpreted in the model of UP as
distribution function of the smeared mass of unstable particle $R_1$. Convolution structure of Eq.(\ref{3.5})
is caused by the factorization of the decay rate $\Gamma(R\to x R_1\to x, all)$.

From Eqs.(\ref{3.4}) and (\ref{3.5}) it follows:
\begin{align}
 &\sigma(ab\rightarrow R\rightarrow x R_1)=\notag\\&\frac{16\pi
 k_R}{k_a k_b\bar{\lambda}^2(m_a,m_b;\sqrt{s})}\frac{\Gamma^{ab}_R(s)}{|P_R(s)|^2}
 \int_{q^2_1}^{q^2_2}\Gamma(R(s)\rightarrow
 x R_1(q))\rho_{R_1}(q)\,dq^2.
 \label{3.6}
\end{align}
Using the expression for $\rho_{R_1}(q)$, from Eq.(\ref{3.6}) we can get the cross-section of exclusive process,
for example $ab\to R\to R_1 x\to cdx$. To this effect we represent $\Gamma ^{tot}_{R_1}(q)$ in the form:
\begin{equation}
 \Gamma^{tot}_{R_1}(q)=\sum_{X_1}\Gamma^{X_1}_{R_1}(q);\,\,\,
 \Gamma^{cd}_{R_1}(q)=\Gamma(R_1(q)\to cd)\,.
 \label{3.7}
\end{equation}
As a result, from (\ref{3.6}) and (\ref{3.7}) we get:
\begin{align}
 &\sigma(ab\rightarrow R\rightarrow xcd)=\notag\\
 &\frac{16k_R}{k_a k_b\bar{\lambda}^2(m_a,m_b;\sqrt{s})}\frac{\Gamma^{ab}_R(s)}{|P_R(s)|^2}
 \int_{q^2_1}^{q^2_2}\Gamma(R(s)\rightarrow
 x R_1(q))\frac{q \Gamma^{cd}_{R_1}(q)}{|P_{R_1}(q)|^2}\,dq^2.
 \label{3.8}
\end{align}
It should be noted that, in analogy with the decay processes, the factorization effectively reduces the number
of independent kinematical variables in the scattering processes too. In the standard approach for the process $2\to 3$ the number of such variables $N=3n-4=5$ \cite{20}, while the approach suggested gives $N=1$.

The processes of scattering with one unstable particle or resonance and one quasistable particle in the final state were discussed in \cite{8a,8b}. Such processes, called in \cite{8a} as pseudoresonances, exhibit themselves as peak in cross-section in analogy with ordinary resonance. However, they are not caused by the pole of $S$-matrix, rather by nonelastic channels \cite{8b}.

Similar structure arises in the case $R\rightarrow R_1R_2$, i.e. when there are two UP in the final state, which have
two-particle decay channels (semi-analytical approach \cite{12}-\cite{14}). Thus, the
model gives a convenient instrument to describe two-particle
scattering accompanied by complicated decay-chain processes. However,
we have checked by direct calculations only two types of processes
- the decay of type $a\rightarrow Rx\rightarrow bcx$ and
the scattering of type $ab\rightarrow R\rightarrow cd$. The more
complicated processes, such as decay $a\rightarrow
R_1 R_2\rightarrow cdef$ and scattering $ab\rightarrow
R\rightarrow R_1 R_2\rightarrow cdef$, will be the subject of
the next paper.

\section{Phenomenology of the factorization method}
In this section, we consider some methodological and phenomenological aspects
of factorization.
The model factorization of a decay width and cross-section of the
processes with UP in an intermediate state was established by
straightforward calculations at tree level. Note that these
calculations in the effective theory of UP \cite{6a,7} account for
some loop contributions. The vertex and self-energy type corrections can be
included into $\Gamma_R(s)$ and $P_R(s)$ respectively. These corrections do
not breakdown a factorization, but the interaction between initial
and final states does. However, such an interaction has no clear
and explicit status in perturbation theory due to UP (or
resonance) is not a perturbative object in the resonance
neighborhood \cite{6a}. As it was noted in Ref.\cite{14}, such
non-factorable corrections give small contribution to the processes
$e^{+}e^{-}\to ZZ, WW, 4f$ in the resonance range.

Now, we consider another aspect of factorization effect, namely, the
determination of dressed propagator of UP. Factorization of decay
width and cross-section does not depend on the structure of
propagator's denominator $P_R(q)$, but crucially depends on the
structure of its numerator in the case of vector and spinor UP. As
it was verified by direct calculations, the factorization always
takes place in the case of scalar UP. The expressions
$\eta_{\mu\nu}(m_R) =g_{\mu\nu}-q_{\mu}q_{\nu}/m^2_R$ and
$\hat{\eta}(m_R)=\hat{q}+m_R$ for vector and spinor UP,
respectively, do not lead to exact factorization. But the expressions
$\eta_{\mu\nu}(q) =g_{\mu\nu}-q_{\mu}q_{\nu}/q^2$ and
$\hat{\eta}(q)=\hat{q}+q$ strictly lead to factorization for any
kinds of particles. It should be noted that the definition
of the functions $\eta_{\mu\nu}(q)$ and $\hat{\eta}(q)$ is not related with
the choice of the gauge, because effective theory of UP \cite{6a} is
not the gauge theory. The choice of $q$ instead of $m_R$ in the
$\eta_{\mu\nu}$ and $\hat{\eta}$ may seems contradict to the
equation of motion for vector and spinor UP. However, this
statement is valid for the stable particle with fixed mass. In the
case of UP the question arises what the mass participates in
equation of motion - pole mass or one of the renormalized
mass? An account of uncertainty relation by smearing of
mass intensifies the question. There is no unique and strict
determination of dressed propagator structure for vector and
spinor UP due to the specific nature of renormalization in these cases
\cite{7}. The situation is more complicated and involved in the
case of hadron resonance. So, the functions $\eta_{\mu\nu}$ and
$\hat{\eta}$ have rather phenomenological (or model) than
theoretical status. The model of UP \cite{6a} defines these
functions as $\eta_{\mu\nu}(q)$ and $\hat{\eta}(q)$, which
describe the dressed propagators of UP in the resonance
neighborhood.

Further, we briefly analyze the phenomenological aspect of
the factorization effect. Universal convolution formula for a decay rate
was widely used in the so-called convolution method (CM). This method introduces the
factorization in a phenomenological way. It was applied for the description
of the near-threshold decays of $t$ quark \cite{9,10} and non-leptonic decays of hadrons \cite{15}-\cite{18}.
The decay rates of the near-threshold decays $t\rightarrow bWZ,cWW,cZZ$ were
calculated within the framework of CM
and DCM (decay-chain method) in Refs.\cite{9,10}. The contributions of FWE lead to
the substantial enhancement of the decay rates, in particular, of
$B(t\rightarrow bWZ)$ and $B(t\rightarrow cZZ)$. For instance, the
branchings without ($B$) and with an accounting of FWE ($\bar{B}$)
in the case of decay $t\rightarrow bWZ$ differ by an order of
magnitude \cite{9}: $B(t\rightarrow bWZ)\sim
10^{-7}$ and  $\bar{B}(t\rightarrow bWZ)\sim 10^{-6}$.
The description of FWE in hadron decays of type $H\rightarrow H_1 H_2$ is
directly follows from the approach suggested, when $H_1$ and (or) $H_2$ are the
hadrons with a large width. The contribution of FWE to decay
rates of the decays $B^0\rightarrow D^-\rho^+$, $B^0\rightarrow
D^-a^+_1$ and $\Lambda^0_b\rightarrow \Lambda^+_c\rho^-$,
$\Lambda^0_b\rightarrow \Lambda^+_c a^-_1$ were considered in
Refs.\cite{15}-\cite{18} within the framework of CM. The result of
calculations reveals that the contributions of FWE are large (from
20 to 40 percent) and its account improves the conformity of the
experimental data and theoretical predictions. In the work \cite{6a}, we have reanalyzed the
decays $B^0\rightarrow D^-\rho^+$ and $\phi(1020)\to K\bar{K}$ within the frame
of the model considered and significantly improve
the correspondence between the experimental data and theoretical predictions.

Now, we analyze the phenomenology of the factorization in the processes of scattering.
In the low-energy experiments of type
$e^+e^-\rightarrow \rho, \omega...\rightarrow \pi^+\pi^-, ...$ we
can not distinguish propagators $\eta_{\mu\nu}(m_R)$ and
$\eta_{\mu\nu}(q)$ even for the wide resonance. This is due to
the equality $\bar{e}^-(p_1)(\hat{p}_1+\hat{p}_2)e^-(p_2)=0$, when the
functions $\eta_{\mu\nu}$ reduce to $g_{\mu\nu}$ in both cases.
In the high-energy experiments of type $e^+e^-\rightarrow
Z\rightarrow \bar{f}f$, where $f$ is quark or lepton (we neglect
$\gamma-Z$ interference), the transverse part of amplitude is
\begin{equation}
 M_q\sim\bar{e}^-(p_1)\hat{q}(c_e-\gamma_5)e^-(p_2)\bar{f}^+(k_1)\hat{q}
 (c_f-\gamma_5)f^+(k_2),
\label{E:4.1}
\end{equation}
where $q=p_1+p_2=k_1+k_2$. From Eq.(\ref{E:4.1}) with the help of
the Dirac equations in momentum representation it follows
\begin{equation}
 M_q\sim m_em_f \bar{e}^-(p_1)\gamma_5
 e^-(p_2)\bar{f}^+(k_1)\gamma_5f^+(k_2).
 \label{E:4.2}
\end{equation}
As a result, we get the terms $m_e m_f/q^2$ and $m_e m_f/m^2_Z$ for $\eta(q)$ and $\eta(m_R)$,
respectively. The  difference of these values is an order of $(m_e m_f/m^2_Z)\cdot(m_Z-q)/m_Z$
at $q^2\sim m^2_Z$. Thus, the distinction between the structure of two type
of the expressions $\eta_{\mu\nu}$ is negligible in a wide range of energy.
We always can evaluate this deviation, that is the approach suggested gives us a simple analytical analog of NWA.
This approach was applied also to the near-threshold $Z$-pair
production \cite{14} in the process $e^+e^-\to ZZ$, where the model polarization matrix (\ref{2.3}) was used.
An accordance with the experiment and Monte-Carlo simulation was demonstrated. From this result, it follows that the contribution of the non-factorable corrections is small at the resonance energy.

The structure of $\hat{\eta}$ can be studied in the process of type
$VF\rightarrow R\rightarrow V^{'}F^{'}$, where $V$ and $F$ are vector and fermion field, $R$ is, for instance, baryon resonance with a large
width. In this case, the difference between $\hat{\eta}(m_R)$ and
$\hat{\eta}(q)$ is characterized by the value $\sim\Gamma_R/m_R$
at peak region, and this problem demands more detailed
consideration.

From this brief analysis it follows that the method of factorization is a simple analytical analog of the narrow-width approximation (NWA, which contains five critical assumptions \cite{5}). Instead, we use the structure of propagators' numerators $\eta(q)$, which follows from usual ones under a simple transformation $m_R \to q$, and one assumption: there is no significant interference with non-resonant processes (fifth assumption of NWA). The rest assumptions of NWA can be derived from the first our point, where some of them are not obligatory in the special cases. The method leads to factorization in the basic type of processes - decay-chain processes (universal convolution formula (\ref{2.7})) and scattering ones (universal factorized formula (\ref{2.11})). Combining these two results, we get a simple and strict algorithm of analytical description of the complicated chain processes.

\section{Conclusion}

The factorization method gives us a convenient semianalytical way
to describe the three-particle decays and two-particle scattering
processes. This effect significantly simplifies calculations and
gives compact universal formulae for the decay rate and
cross-section.

In this work, we have shown that the factorization always is
valid when scalar UP is in the intermediate state. In the case of
vector or spinor intermediate states, the factorization takes
place when the specific propagators are used for these states. These
propagators are derived in the model of UP with a random (smeared) mass.
They negligibly differ from the traditional propagators at peak area and
follow from the smearing of mass in accordance
with an uncertainty relation. Our method makes it
possible significantly simplify the description of the complicated
decay-chain and scattering processes. It is some analytical analog of NWA
and gives a simple and strict algorithm for calculations. This approach can be
treated also as a convenient approximation to the traditional one, which always is valid in the
resonance range, where non-resonance contribution is small.

We have performed also a short methodological and phenomenological
analysis of the approach under discussion. It was shown, that in
the process $e^+e^-\rightarrow f\bar{f}$ the difference between
two forms of propagators is negligible in a wide range of energy. It can be significant in the processes with baryon resonance in an intermediate state, but in this case we should perform an additional analysis.

\appendix
\section{Appendix 1}

In this section, we represent the model formalism we need to construct the propagators for the vector and spinor fields (for the detail see \cite{6,6a}). The structure of these propagators lead to the factorization effect in the processes with the participation of UP in the intermediate state. The model field wave function, which describes UP, is
\begin{equation}\label{A.1}
 \Phi_a(x)=\int\Phi_a(x,\mu)\omega(\mu)d\mu,
\end{equation}
where $\Phi_a(x,\mu)$ is spectral component, which defines a particle
with a fixed mass squared $m^2=\mu$ in the stable particle
approximation (SPA):
\begin{equation}\label{A.1a}
 \Phi_{\alpha}(x,\mu)=\frac{1}{(2\pi)^{3/2}}\int\Phi_{\alpha}(k)
 \delta(k^2-\mu)e^{ikx}\,dk.
\end{equation}
The weight function $\omega(\mu)$ is
formed by the self-energy type interactions of UP with vacuum fluctuations
and decay products. This function describes the smeared (fuzzed)
mass-shell of UP.

The model Lagrangian, which determines a "free" (effective) unstable field
$\Phi(x)$, has the convolution form:
\begin{equation}\label{A.2}
 L(\Phi(x))=\int L(\Phi(x,\mu))|\omega(\mu)|^2\,d\mu\,.
\end{equation}
In Eq.(\ref{A.2}) $L(\Phi(x,\mu))$ is the standard Lagrangian,
which describes model "free" field component $\Phi(x,\mu)$ in the stable
particle approximation ($m^2=\mu$).

From Eq.(\ref{A.2}) and prescription
$\partial\Phi(x,\mu)/\partial\Phi(x,\mu^{'})=\delta(\mu-\mu^{'})$
it follows the Klein-Gordon equation for the spectral component of
the unstable field:
\begin{equation}\label{A.3}
 (\square-\mu)\Phi_{\alpha}(x,\mu)=0.
\end{equation}
As a result, we get the  standard representation (\ref{A.1a}) of the spectral component $\Phi_{\alpha}(x,\mu)$ with a fixed mass parameter $\mu$. All standard definitions, relations and frequency expansion take
place for $\Phi_{\alpha}(k,\mu)$, however, the relation $k^0_{\mu}=\sqrt{{\bf k}^2+\mu}$ defines the smeared (fuzzy) mass-shell due to a random nature of the mass parameter $\mu$. In analogy with (\ref{A.3}) one can get the Dirac equation $(i\hat{\partial}+\sqrt{\mu})\Psi(x,\mu)=0$ for fermion spectral component. The convolution (diagonal) representation of the "free" Lagrangian (\ref{A.2}) has an
assumption (or approximation?) that the states with different $\mu$
do not interact in the approximation of the model "free" fields.

The expressions (\ref{A.1})--(\ref{A.3}) define the model
"free" unstable field as some effective field. As it was mentioned
above, this field is formed by an interaction of "bare" UP with
the vacuum fluctuations and decay products, that is includes
the self-energy contribution. This interaction leads to the spreading (smearing) of mass, described by
the function $\omega(\mu)$ or $\rho(\mu)=|\omega(\mu)|^2$. Thus, we go from the
distribution $\rho^{st}(\mu)=\delta(\mu-M^2)$ for "bare" particles
to some smooth density function $\rho(\mu)=|\omega(\mu)|^2$ with
mean value $\bar{\mu}\approx M^2$ and mean square deviation
$\sigma_{\mu}\approx \Gamma$. So, the UP is characterized by the
weight function $\omega(\mu)$ or probability density $\rho(\mu)$
with parameters $M$ and $\Gamma$ (or real and imaginary parts of a
pole).

The commutative relations for the model operators have an additional
$\delta$-function:
\begin{equation}\label{A.4}
 [\dot{\Phi}^{-}_{\alpha}({\bf k},\mu),\,\Phi^{+}_{\beta}({\bf q},\mu^{'})]_{\pm}
 =\delta(\mu-\mu^{'}) \delta({\bf k}-{\bf q})\delta_{\alpha\beta},
\end{equation}
where subscripts $\pm$ correspond to the fermion and boson fields.
The presence of $\delta(\mu-\mu^{'})$ in Eq.(\ref{A.4}) means an
assumption - the acts of creation and annihilation of the
particles with different $\mu$ (the random mass squared) do not
interfere. Thus, the parameter $\mu$ has the status of physically
distinguishable value of a random $m^2$. This assumption is
naturally related with a diagonal form of Eqs.(\ref{A.2}) and
(\ref{A.3}). By integrating both sides of Eq.(\ref{A.4}) with weights
$\omega^{*}(\mu)\omega(\mu^{'})$ one can get the standard
commutative relations
\begin{equation}\label{A.5}
 [\dot{\Phi}^-_{\alpha}({\bf k}),\Phi^+_\beta({\bf q})]_{\pm}=\delta({\bf k}-{\bf q})
 \delta_{\alpha\beta}\,,
\end{equation}
where $\Phi^{\pm}_{\alpha}({\bf k})$ is full operator field
function in the momentum representation:
\begin{equation}\label{A.6}
 \Phi^{\pm}_{\alpha}({\bf k})=\int\Phi^{\pm}_{\alpha}({\bf k},\mu)\omega(\mu)d\mu\,.
\end{equation}

The amplitude for the transition $\Phi\rightarrow R\phi_1$, where $R$ is scalar UP
with a large width, has the form
\begin{equation}\label{A.7}
 A(k,\mu)=\omega(\mu)A^{st}(k,\mu)\,,
\end{equation}
where $A^{st}(k,\mu)$ is the amplitude in a stable particle
approximation. This amplitude is calculated in the standard way
and can include the higher corrections.
The differential (on ${\bf k}$) probability is
\begin{equation}\label{A.8}
 d\Gamma(k)=\int d\Gamma^{st}(k,\mu)\rho(\mu)d\mu\,.
\end{equation}
In Eq.(\ref{A.8}) the differential probability
$d\Gamma^{st}(k,\mu)$ is defined in the standard way (stable
particle approximation):
\begin{equation}\label{A.9}
 d\Gamma^{st}(k,\mu)=\frac{1}{2\pi}\delta(k_{\Phi}-k_{R}-k_1)|A^{st}
 (k,\mu)|^2d{\bf k}_{\phi}d{\bf k}_1\,,
\end{equation}
where $k=(k_{\Phi},k_{R},k_1)$ denotes the 4-momenta of
particles. From Eqs.(\ref{A.8}) and (\ref{A.9}) it directly
follows the well-known convolution formula (CF) for a decay rate
\begin{equation}\label{A.10}
 \Gamma(m_{\Phi},m_1)=\int_{\mu_1}^{\mu_2}\Gamma^{st}(m_{\Phi},m_1;\mu)\rho(\mu)d\mu\,,
\end{equation}
where $\rho(\mu)=|\omega(\mu)|^2$, $\mu_1$ and $\mu_2$ are the
threshold and maximal invariant mass squared of an unstable
particle.
If there are two UP with large widths in a final state of decay
$\Phi\rightarrow R_1 R_2$, then in analogy with the previous
case one can get the double convolution formula:
\begin{equation}\label{A.11}
 \Gamma(m_{\Phi})=\int\int\Gamma^{st}(m_{\Phi};\mu_1,\mu_2)\rho_1(\mu_1)\rho_2(\mu_2)d\mu_1
 d\mu_2\,.
\end{equation}
The derivation of CF for the cases, when there is a vector or
spinor UP in the final state, can be done in analogy with the case
of scalar UP. However, in Eqs.(\ref{A.7}) and (\ref{A.9}) one should take into account the polarization vector
$e_m(q)$ or spinor $u^{\nu,\pm}_{\alpha}(q)$, where momentum $q$ is on fuzzy
mass-shell. In this case, we derive the polarization matrixes in full analogy with standard derivation, but taking into consideration modified Klein-Gordan (\ref{A.3}) and Dirac equations. As a result, we get the polarization matrix with $m^2=\mu$, that is on smeared mass-shell. In the case of vector UP in the final state
we have:
\begin{equation}\label{A.12}
 \sum_{e} e_m(q)e^{*}_n(q)=-g_{mn}+q_mq_n/\mu\,,
\end{equation}
where $q^0_{\mu}=\sqrt{{\bf q}^2+\mu}$ and the summation over polarization is implied.
In the case of spinor UP in the final state:
\begin{equation}\label{A.13}
 \sum_{\nu} u^{\nu,\pm}_{\alpha}(q)\bar{u}^{\nu,\mp}_{\beta}(q)=\frac{1}{2q^0_{\mu}}
 (\hat{q}\mp\sqrt{\mu})_{\alpha\beta}\,,
\end{equation}
The same relations take place for the initial states,
however one have to average over the polarizations.

Now, we consider the structure of the model propagators.
With the help of the traditional method, one can get from
Eqs.(\ref{A.1}), (\ref{A.4}) and (\ref{A.6}) the expression for
the unstable scalar Green function \cite{6}:
\begin{equation}\label{A.14}
 \langle 0|T(\phi(x),\phi(y))|0\rangle\equiv D(x-y)=\int
 D(x-y,\mu)\rho(\mu)d\mu\,.
\end{equation}
In Eq.(\ref{A.14}) $D(x,\mu)$ is a standard scalar Green function,
which describes UP in an intermediate state with a fixed $m^2=\mu$:
\begin{equation}\label{A.15}
 D(x,\mu)=\frac{i}{(2\pi)^4}\int\frac{e^{-ikx}}{k^2-\mu+i\epsilon}dk\,.
\end{equation}
The right-hand side of Eq.(\ref{A.14}) is the Lehmann-like
spectral (on $\mu$) representation of the scalar Green function. Taking into
account the relation between scalar and vector Green functions, we
can get the Green function of the vector unstable field in the
form:
\begin{align}\label{A.16}
 D_{mn}(x,\mu)=&-(g_{mn}+\frac{1}{\mu}\frac{\partial^2}{\partial
 x^n\partial x^m})D(x,\mu)\notag\\
 =&\frac{-i}{(2\pi)^4}\int\frac{g_{mn}-k_m
 k_n/\mu}{k^2-\mu+i\epsilon}e^{-ikx}dk\,.
\end{align}
Analogously, the Green function of the spinor unstable field is
\begin{equation}\label{A.17}
 \Hat{D}(x,\mu)=(i\hat{\partial}+\sqrt{\mu})D(x,\mu)=\frac{i}{(2\pi)^4}
 \int\frac{\hat{k}+\sqrt{\mu}}{k^2-\mu+i\epsilon}e^{-ikx}dk\,,
\end{equation}
where $\hat{k}=k_i\gamma^i$. These Green functions in momentum
representation have a convolution form:
\begin{equation}\label{A.18}
 D_{mn}(k)=-i\int \frac{g_{mn}-k_mk_n/\mu}{k^2-\mu+i\epsilon} \rho(\mu)d\mu\,,
\end{equation}
and
\begin{equation}\label{A.19}
 \hat{D}(k)=i\int \frac{\hat{k}+k}{k^2-\mu+i\epsilon} \rho(\mu)d\mu\,,
\end{equation}
To construct the model propagators in explicit form we need to define the probability density
$\rho(\mu)$. Here, we represent the definition of $\rho(\mu)$ from the
matching the model propagators to the standard dressed ones \cite{6,6a}.
We associate the model propagator of scalar unstable
field (\ref{A.14}, \ref{A.15}) in a momentum representation with the standard dressed one:
\begin{equation}\label{A.20}
 \int\frac{\rho(\mu)d\mu}{k^2-\mu+i\epsilon}\longleftrightarrow
 \frac{1}{k^2-m^2_0-\Pi(k^2)}\,,
\end{equation}
where $\Pi(k^2)$ is the conventional polarization operator ( or
self-energy) of scalar field. With the help of the analytic continuation method it was shown in \cite{6}, that the conformity (\ref{A.20}) leads to the definition:
\begin{equation}\label{A.21}
 \rho(\mu)=\frac{1}{\pi}\,\frac{Im\Pi(\mu)}{[\mu-m^2(\mu)]^2+[Im\Pi(\mu)]^2}\,,
\end{equation}
where $m^2(\mu)=m^2_0+Re\Pi(k^2)$.
The expression (\ref{A.21}) for $\rho(k^2)$ in the Breit-Wigner
approximation is usually exploited within the framework of the
convolution method.

Inserting $\rho(\mu)$ into the
model propagator (\ref{A.18}) for vector unstable field
leads to the result \cite{6a}:
\begin{equation}\label{A.22}
 D_{mn}(k)=-i\frac{g_{mn}-k_m
 k_n/k^2}{k^2-m^2(k^2)-iIm\Pi(k^2)}\,.
\end{equation}
An analogous procedure with Eq.(\ref{A.19}) and the change $Im\Pi(\mu)\to \sqrt{\mu}Im\Sigma(\mu)$
leads to the definition:
\begin{equation}\label{A.23}
 \hat{D}(k)=i\frac{\hat{k}+k}{k^2-m^2(k^2)-ik\Sigma(k^2)}\,,
\end{equation}
Eqs. (\ref{A.20})-- (\ref{A.23}) establish the correspondence between the model under consideration
and some effective theory of UP in the framework of traditional QFT
approach. This effective theory has a close analogy
with the traditional description of UP in the intermediate state
as a special case of the approach discussed. The most important
features of the effective theory, constructed in such a way, are
the factorization and convolution effects (see Section 2 and Appendix B).
These effects arise due to the specific structure of the propagator's
numerators $\eta_{\mu\nu}(k)=g_{mn}-k_m k_n/k^2$ and $\hat{\eta}(k)=\hat{k}+k$
($M\to k$, smearing of the mass-shell).

\section{Appendix 2}

In this section, we consider convenient and simple method of calculation of three-particle
decay rate and two-particle cross-section. This method is based on the model of UP with a smeared mass,
where the expressions for polarization matrixes (\ref{A.12}), (\ref{A.13}) and propagators (\ref{A.20})--(\ref{A.23}) are constructed.  The vertexes are defined by
the Lagrangian in the simplest standard form:
\begin{align}
 L_k=&g\phi\phi_1\phi_2;\,\,\,g\phi\bar{\psi}_1 \psi_2;\,\,\,g\phi
 V_{1\mu}V^{\mu}_2;\,\,\,gV_{\mu}(\phi_1^{,\mu}\phi_2-\phi_2^{,\mu}\phi_1);\,\,\,
 gV_{\mu}\bar{\psi}_1\gamma^{\mu}(c_V+c_A\gamma_5)\psi_2;\notag\\&gV_{1\mu}V_{2\nu}V_{\alpha}
 [g^{\mu\nu}(p_2-p_1)^{\alpha}+g^{\mu\alpha}(2p_1+p_2)^{\nu}-g^{\nu\alpha}(p_1+2p_2)^{\mu}].
 \label{B.1}
\end{align}
In the expressions (\ref{B.1}) $\phi, V$ and $\psi$ are the
scalar, vector and spinor fields, respectively, $p_1$ and $p_2$
are the momenta of particles.
It is convenient to employ the universal expressions for widths
$\Gamma(R\rightarrow ab)$ or $\Gamma(a\rightarrow Rb)$ in a
stable particle approximation:
\begin{equation}
 \Gamma_i(R\rightarrow
 ab)=\frac{g^2}{8\pi}\bar{\lambda}(m_a,m_b;m_R)f_i(m_a,m_b;m_R),
\label{B.2}
\end{equation}
where $m_R^2=q^2=(p_1+p_2)^2$ and:
\begin{equation}
 \bar{\lambda}(m_a,m_b;m_R)=[1-2\frac{m^2_a+m^2_b}{m^2_R}+\frac{(m^2_a-m^2_b)^2}{m^4_R}]^{1/2}.
\label{B.3}
\end{equation}
The same expressions and relations are in order for the width
$\Gamma(a\rightarrow Rb)$. The functions $f_i(m_a,m_b;m_R)$ are
defined by the corresponding vertexes. If these vertexes are
described by Eqs.(\ref{B.1}), then the functions $f_i$ (further
we omit the arguments) in tree approximation are defined by the following
expressions:
\begin{align}
 &\phi\rightarrow
 \phi_1\phi_2,\,\,f_1=\frac{1}{2m_{\phi}};\,\,\,\,\phi\rightarrow V_1V_2,\,\,
 f_2=\frac{1}{m_{\phi}}[1+\frac{(m^2_{\phi}-m^2_1-m^2_2)^2}{8m^2_1m^2_2}];\notag\\
 &\phi\rightarrow
 \bar{\psi}_1\psi_2,\,\,f_3=m_{\phi}[1-\frac{(m_1+m_2)^2}{m^2_{\phi}}];\,\,\,
 \phi\rightarrow\phi_1V,\,\,f_4=\frac{m^3_{\phi}}{2m^2_V}\bar{\lambda}^2(m_1,m_V;m_{\phi});\notag\\
 &V\rightarrow\phi_1\phi_2,\,\,f_5=\frac{m_V}{6}\bar{\lambda}^2(m_1,m_2;m_V);\,\,\,\,
 V\rightarrow V_1\phi,\,\,f_6=\frac{1}{3m_V}[1+\notag\\ &\,\,\,+\frac{(m^2_V+m^2_1-m^2_{\phi})^2}{8m^2_Vm^2_1}];\notag\\
 &V\rightarrow\bar{\psi}_1\psi_2,\,\,f_7=\frac{2}{3}m_V\{c_{+}[1-\frac{m^2_1+m^2_2}{2m^2_V}-
 \frac{(m^2_1+m^2_2)^2}{2m^4_V}]+3c_{-}\frac{m_1m_2}{m^2_V}\};\notag\\
 &V\rightarrow V_1V_2,\,\,f_8=\frac{m^5_V}{24m^2_1m^2_2}[1+8(\mu_1+\mu_2)-2(9\mu^2_1+16\mu_1\mu_2+9\mu^2_2)+
 8(\mu^3_1-\notag\\ &4\mu^2_1\mu_2
 -4\mu_1\mu^2_2+\mu^3_2)+\mu^4_1+8\mu^3_1\mu_2-18\mu^2_1\mu^2_2+
 8\mu_1\mu^3_2+\mu^4_2],\,\,\mu_{1,2}=m^2_{1,2}/m^2_V;\notag\\
 &\psi\rightarrow
 \phi\psi_1,\,\,f_9=\frac{m_{\psi}}{2}(1+2\frac{m_1}{m_{\psi}}+\frac{m^2_1-m^2_{\phi}}{m^2_{\psi}});\notag\\
 &\psi\rightarrow
 V\psi_1,\,\,f_{10}=m_{\psi}\{c_{+}[\frac{(m^2_{\psi}-m^2_1)^2}{2m^2_{\psi}m^2_V}+\frac{m^2_{\psi}+m^2_1
 -2m^2_V}{2m^2_{\psi}}]-3c_{-}\frac{m_1}{m_{\psi}}\};\notag\\
 &c_{+}=c^2_V+c^2_A,\,\,\,c_{-}+c^2_V-c^2_A\,.
\label{B.4}
\end{align}

Using the expressions (\ref{B.1})--(\ref{B.4}) we can represent
$\Gamma(\Phi\rightarrow\phi_1\phi_2\phi_3)$, that is the width of the process $\Phi\to \phi_1 R\to \phi_1 \phi_2 \phi_3$, in a compact and universal form for all types of decay channels. Here we shortly
describe the method of calculation the width $\Gamma(\Phi\rightarrow\phi_1\phi_2\phi_3)$.
This value always can be written as:
\begin{equation}\label{B.5}
 \Gamma(\Phi\rightarrow\phi_1\phi_2\phi_3)=\frac{k}{p^0}\int J(\vert M(k_i,m_i)\vert^2)\frac{d{\bf k}_1}{k^0_1}\,,
\end{equation}
where $M(k_i,m_i)$ is an amplitude, $p$ and $k_i$ are momentum of
$\Phi$ and $\phi_i$, $k$ is some numerical factor, and
\begin{equation}\label{B.6}
 J(\vert M\vert^2)=\int \vert M\vert^2\delta(p-k_1-k_2-k_3)\frac{d{\bf k}_2d{\bf k}_3}
 {k^0_2k^0_3}.
\end{equation}
The integral $J(\vert M\vert^2)$ is easily calculated in ${\bf q}={\bf k}_2+{\bf k}_3=0$ frame of reference.
As a result, we have the non-covariant expression
\begin{equation}\label{B.7}
 J(\vert M\vert^2)\,\longrightarrow\,f(q^0,q^0 p^0,q^0 k^0_3,{\bf p}^2,...).
\end{equation}
This expression can be always reconstructed to covariant form by the transition (we use ${\bf q}=0$):
\begin{equation}\label{B.8}
 q^0\rightarrow q=\sqrt{(qq)},\,q^0p^0\rightarrow (qp),\,q^0k^0_1\rightarrow (qk_1),\,
 \bar{p}^2=(p^0)^2-m^2\rightarrow (pq)^2/q^2-m^2,...
\end{equation}
Then we pass to the ${\bf p}=0$ frame of reference and change the
variable in Eq. (\ref{B.5}) according to
\begin{equation}\label{B.9}
 \frac{d{\bf k}_1}{k^0_1}=-\frac{1}{2m}|{\bf k}_1| dq^2 d\Omega=-\frac{1}{4}
 \tilde{\lambda}(q,m_1;m)dq^2d\Omega.
\end{equation}

Using this simple method and expressions for the propagators (\ref{A.22}), (\ref{A.23}), we have got
by tedious but straightforward calculations the general expression for
$\Gamma(\Phi\to\phi_1\phi_2\phi_3)=\Gamma(\Phi\to\phi_1 R\to\phi_1\phi_2\phi_3)$, where $\Phi, R$ and $\phi_k$ are particles of all possible type:
\begin{equation}\label{B.10}
 \Gamma_{\alpha\beta}(\Phi\rightarrow\phi_1\phi_i\phi_k)=\frac{g^2_1g^2_2}{2^6\pi^3}
 \int_{q^2_1}^{q^2_2}\tilde{\lambda}(q,m_1;m)f_{\alpha}(q,m_1;m)\tilde{\lambda}
 (m_i,m_k;q)f_{\beta}(m_i,m_k;q)\frac{qdq^2}
 {\vert P_{R}(q)\vert^2}\,,
\end{equation}
where $q_1=m_i+m_k$ and $q_2=m-m_1$. From Eqs. (\ref{B.10}) and
(\ref{B.2}) it follows:
\begin{equation}\label{B.11}
 \Gamma_{\alpha\beta}(\Phi\rightarrow \phi_1\phi_i\phi_k)=\int_{q^2_1}^{q^2_2}
 \Gamma_{\alpha}(\Phi\rightarrow\phi_1 R(q))\frac{q\Gamma_{\beta}(R(q)
 \rightarrow\phi_i \phi_k)}{\pi\vert P_{R}(q)\vert^2}\,dq^2\,,
\end{equation}
where $\alpha$ and $\beta$ denote the type of decay in (\ref{B.2})--(\ref{B.4}).
In the approximation
\begin{equation}\label{B.12}
 \Gamma(\Phi\rightarrow\phi_1 R)=\sum_{i,k}\Gamma(\Phi\rightarrow\phi_1\phi_i
 \phi_k)
\end{equation}
we get the well-known convolution formula
\begin{equation}\label{B.13}
 \Gamma(\Phi\rightarrow\phi_1 R)=\int_{q^2_1}^{q^2_2}\Gamma(\Phi\rightarrow
 \phi_1 R(q))\rho_{R}(q)dq^2\,,
\end{equation}
where
\begin{equation}\label{B.14}
 \rho_{R}(q)=\frac{q}{\pi\vert P_{R}(q)\vert^2}\sum_{i,k}\Gamma(R(q)\rightarrow \phi_i\phi_k).
\end{equation}
The same result can be received for many-particle decay channels of UP $\Phi\rightarrow
\phi_1\phi_2\phi_3\phi_4...$\,. For example, let us consider the decay chain $\Phi
\rightarrow\phi_1 R\rightarrow\phi_1\phi_2\phi_3\phi_4$, where $\phi_k$ are the
scalar fields. Then for the simplest contact interaction we have:
\begin{equation}\label{B.15}
 \Gamma_{\Phi}=\frac{g^2_1g^2_2}{2^{13}\pi^8 p^0}\int\frac{d{\bf k}_1}
 {k^0_1\vert P_{R}(q)\vert^2}\int\int\int\delta(q-k_2-k_3-k_4)\frac
 {d{\bf k}_2d{\bf k}_3d{\bf k}_4}{k^0_2k^0_3k^0_4}\,,
\end{equation}
where $q=p-k_1$. The width of the intermediate decay is
\begin{equation}\label{B.16}
 \Gamma_{R}(q)\equiv \Gamma(R(q)\rightarrow\phi_1\phi_2\phi_3)=
 \frac{g^2_2}{2^9\pi^5q^0}\int\int\int\delta(q-k_2-k_3-k_4)\frac
 {d{\bf k}_2d{\bf k}_3d{\bf k}_4}{k^0_2k^0_3k^0_4}\,.
\end{equation}
From Eqs. (\ref{B.15}), (\ref{B.16}) with the help of (\ref{B.2}) and (\ref{B.9}) we get:
\begin{equation}\label{B.17}
 \Gamma_{\Phi}=\int_{q^2_1}^{q^2_2}dq^2\Gamma_{\Phi}(q)\frac{q\Gamma_{R}(q)}
 {\pi\vert P_{R}(q)\vert^2}\,,
\end{equation}
where $\Gamma_{\Phi}(q)\equiv\Gamma(\Phi\rightarrow\phi_1 R(q))$.
Thus we have illustrated the validity of factorization in the case of scalar UP.
Using the factorable $\vert M\vert^2$, we can get the result
(\ref{B.17}) by direct calculations for others types of particles
$\phi_k$. It should be noted that the factored (\ref{B.11}) and
convolution (\ref{B.13}) structures are valid for any choice of
$P_{R}(q)$.

Now,we consider inelastic scattering of type
$ab\rightarrow R\rightarrow cd$, where $R$ is the UP with a
large width in $s$-channel and $a, b, c, d$ are stable
(quasi-stable) particles of any kind. The vertexes are defined by
the same Lagrangian (\ref{B.1}). In the further calculations it is
convenient to employ the relations,
which take place in the center-of-mass system:
\begin{align}
 &p^0_1=\frac{1}{2}q[1+\frac{m^2_a-m^2_b}{q^2}],\,\,\,
 p^0_2=\frac{1}{2}q[1+\frac{m^2_b-m^2_a}{q^2}],\notag\\
 &(p_1q)=\frac{1}{2}(q^2+m^2_a-m^2_b),\,\,\,
 (p_2q)=\frac{1}{2}(q^2+m^2_b-m^2_a),\notag\\
 &(p_1p_2)=\frac{1}{2}(q^2-m^2_a-m^2_b),\,\,\,|{\bf p}_1|=|{\bf p}_2|=
 \frac{1}{2}q\bar{\lambda}(m_a,m_b;q),
 \label{B.18}
 \end{align}
where $p_1$ and $p_2$ are the momenta of the particles $a$ and $b$.
The analogous relations occur for the momenta $k_1$ and $k_2$
of the particles $c$ and $d$. In Eqs.(\ref{B.18}) the symbol $q$
has different meanings in the expressions $(p_1 q)$, $q=p_1+p_2$
(q is 4-momentum) and in the expression $q[1+f(q)]$, where
$q=\sqrt{(q\cdot q)}$ is a number.

With the help of the relations (\ref{B.1})-(\ref{B.4}), (\ref{B.18}) and using above
discussed expressions for propagators, we have got by tedious but straightforward calculations the universal factorized cross-section for all permissible
combinations of particles $(a,b,R,c,d)$:
\begin{equation}
 \sigma(ab\rightarrow R\rightarrow cd)=\frac{16\pi (2J_R+1)}
 {(2J_a+1)(2J_b+1)(2J_R+1)\bar{\lambda}^2(m_a,m_b;\sqrt{s})}
 \frac{\Gamma^{ab}_R(s)\Gamma^{cd}_R(s)}{|P_R(s)|^2}.
\label{B.19}
\end{equation}
In Eq.(\ref{B.19})  $J_k$ is spin of the particle ($k=a,b,R$),
$s=(p_1+p_2)^2$, $\Gamma^{ab}_R(s)=\Gamma(R(s)\rightarrow ab)$ and
$P_R(s)$ is propagator's denominator of the UP or resonance $R$.
The expressions for $\Gamma^{ab}_R(s)$ and $\Gamma^{cd}_R(s)$
follow from Eqs.(\ref{B.2})-(\ref{B.4}), when squared mass of UP
is $m^2_R=q^2=s$. The factorization of cross-section
does not depend on the definition of $P_R(s)$, which can be
determined in a phenomenological way, in Breit-Wigner or pole form
, etc. The expression (\ref{B.19}) is a natural
generalization of the spin-averaged Breit-Wigner (non-relativistic)
cross-section, defined by the expression (37.51) in Ref.\cite{11}.
Note that the factorization is exact in our approach,
while in the traditional one it occurs as an approximation.

\end{document}